\documentclass[twocolumn,superscriptaddress,secnumarabic,amssymb, nobibnotes, aps, prl]{revtex4-1}
\usepackage{graphicx}
\usepackage{amsfonts}
\usepackage{amssymb}
\usepackage{amsmath}
\usepackage{psfrag}
\usepackage{natbib}
\usepackage{mathtools}
\usepackage{ulem}
\usepackage{comment}

\usepackage[colorlinks=true, linkcolor=blue, urlcolor=blue, citecolor=black]{hyperref}
\usepackage{caption}
\newcommand{\beqa}{\begin{eqnarray}}
\newcommand{\eeqa}{\end{eqnarray}}
\newcommand{\beq}{\begin{equation}}
\newcommand{\eeq}{\end{equation}}

\newcommand{\blue}[1]{\textcolor{blue}{#1}}

\usepackage[justification=justified,singlelinecheck=false]{caption}

\usepackage{ragged2e}

\begin{document}

\title{Creases as elastocapillary gates for autonomous droplet control}

\author{Zixuan Wu}
 \affiliation{Mechanical \& Aerospace Engineering Department, Syracuse University, Syracuse, NY 13244, USA}
\author{Gavin Linton}
 \affiliation{Mechanical \& Aerospace Engineering Department, Syracuse University, Syracuse, NY 13244, USA}
 
\author{Stefan Karpitschka}
 \affiliation{Fachbereich Physik, Universit\"{a}t Konstanz, 78457 Konstanz, Germany}

\author{Anupam Pandey}
\email{apande05@syr.edu}
 \affiliation{Mechanical \& Aerospace Engineering Department, Syracuse University, Syracuse, NY 13244, USA}
 \affiliation{BioInspired Syracuse: Institute for Material and Living Systems, Syracuse University, Syracuse, NY 13244, USA}

\date{\today}

\begin{abstract}
\noindent \textbf{Droplets are the core functional units in microfluidic technologies that aim to integrate computation and reaction on a single platform. Achieving directed transport and control of these droplets typically demands elaborate substrate patterning, modulation of external fields, and real-time feedback. Here we reveal that an engineered pattern of creases on a soft interface autonomously gate and steer droplets through a long-range elastocapillary repulsion, allowing programmable flow of information. Acting as an energy barrier, the crease bars incoming droplets below a critical size,  without making contact. We uncover the multi-scale, repulsive force-distance law describing interactions between a drop and a singular crease. Leveraging this mechanism, we demonstrate passive and active filtration based on droplet size and surface tension, and implement functionalities such as path guidance, tunable hysterons, pulse modulators, and elementary logic operations like adders. This crease-based gating approach thus demonstrates complex in-unit processing capabilities -- typically accessible only through sophisticated surface and fluidic modifications -- offering a multimodal, potentially rewritable strategy for droplet control in interfacial assembly and biochemical assays.}

\end{abstract}

\maketitle

\section{Introduction}
\noindent Droplets and bubbles -- discrete fluid units capable of carrying material and information -- can navigate pre-designed paths, process logic, and enable automated decision-making and reactions in physically intelligent systems \cite{fredkin1982conservative,preston2019digital, prakash2007microfluidic,sitti2021physical}. Digital microfluidics provides the core platform for realizing such capabilities, enabling precise control over fluid motion at small scales and supporting a wide range of applications, from healthcare to environmental monitoring~\cite{faustino2016biomedical}. Achieving programmability on these platforms, however, often demands extensive and irreversible modifications to the chemical \cite{schutzius2015spontaneous,daniel2001fast,bjelobrk2016thermocapillary, chen2016continuous,kusumaatmaja2007controlling}, electrical \cite{abdelgawad2009digital,abdelgawad2008all}, magnetic \cite{vialetto2017magnetic, timonen2013switchable}, or optical \cite{ichimura2000light} properties of the substrate \cite{sun2019surface} or the fluid medium. Moreover, many systems rely on off-board electronics for decision-making and closed-loop feedback \cite{pit2015high,ahn2006dielectrophoretic,jing2015jetting,niu2007real}. These requirements can pose significant barriers to the rapid prototyping, scaling, and reconfiguration of microfluidics~\cite{faustino2016biomedical} in low-resource regions. 

Programmable fluidic systems often establish controls by harnessing nonlinear fluid interactions, which is typically lost at small scales and low speeds. Non-Newtonian fluids \cite{groisman2003microfluidic}, flow mediated bubble-bubble interactions \cite{prakash2007microfluidic}, pressure-flow instabilities \cite{gopinathan2023microfluidic} and capillary-driven mechanics~\cite{martinez2024fluidic,zeng20223d} have been utilized to reintroduce nonlinearity in these low Reynolds number systems. While these strategies exploit nonlinearities within the fluid, soft polymeric substrates offer a novel route -- controlling droplet motion through nonlinear elasto-capillary coupling at the interface. Droplets sense and respond to changes in the mechanical state of the soft substrate they travel on, modulating their speed and direction~\cite{PhysRevLett.118.198002,smith2021droplets, ChaoPRL25}. This mechano-adaptive response opens up opportunities for local, reversible tuning of droplet transport and realizing reconfigurable droplet platforms. 

\begin{figure*}[h!tbp]
    \begin{center}
    \centering
    \includegraphics[width=1\textwidth]{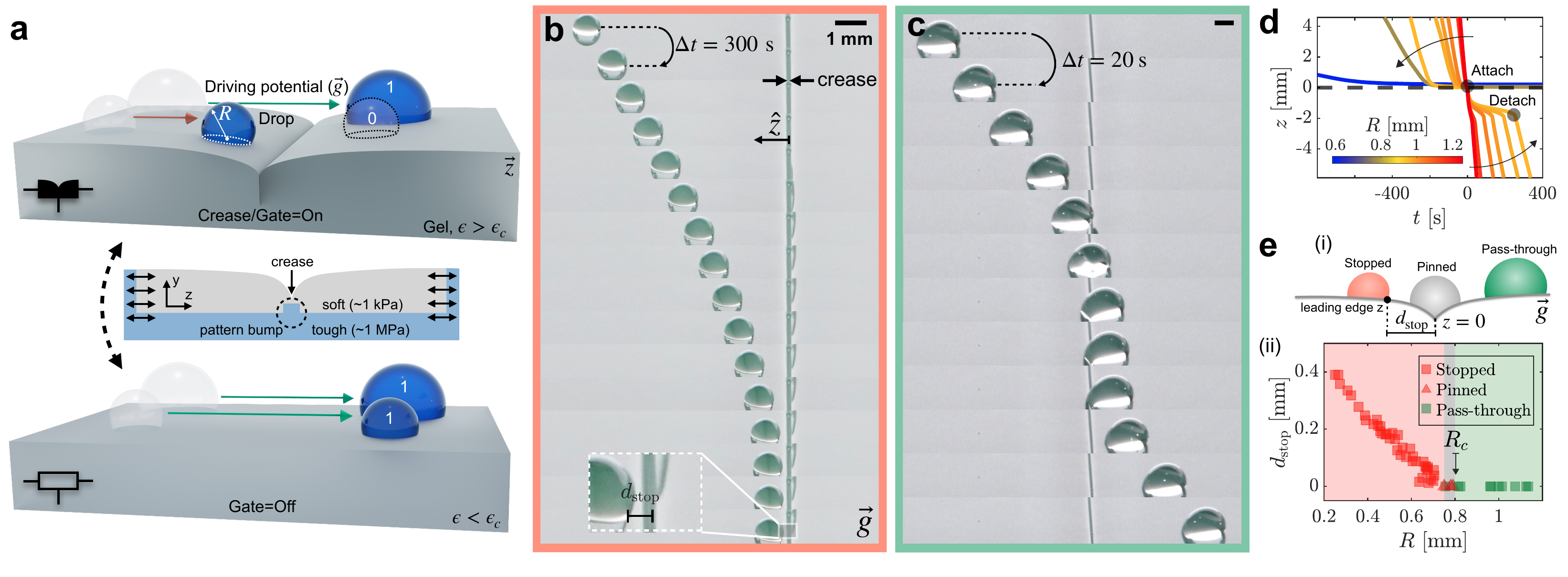}
    \caption{\justifying\textbf{Size-based gating of droplets by a crease on soft elastomer surfaces.} \textbf{a.} Schematics of crease as switchable autonomous gates for interfacial droplets. The crease gate can be toggled between on and off states by varying compressive strain $\epsilon$ above or below the instability threshold $\epsilon_c$ via the dual-layer setup shown in the middle. Symbolic representations of these two gating states are labeled on the left corners of the gel basis. \textbf{b.} Time-lapse images showing a small glycerin droplet, with diameter of 0.8 mm, driven by gravity and halted before reaching a crease. \textbf{c.} A larger glycerin droplet (diameter of 2.2 mm) overcomes the crease and continues its motion. For both these panel the soft substrate is under a compressive strain of 25$\%$. All scale bars are at 1 mm. \textbf{d.} Evolution of the droplet's leading edge with respect to the crease, $z$ over time for different drop radii $R$. \textbf{e.} (i). Schematics of three droplet responses near the the crease: stopped, pinned, pass-through. The critical radius $R_c$ marks the onset of pass-through; droplets with $R<R_c$ either stop or pin, while those with $R>R_c$ cross the barrier. (ii). Stopping distances $d_{\mathrm{stop}}$ versus droplet size, revealing a critical gating radius $R_c\simeq 0.8$ mm.}
    \label{fig1}
    \end{center}
    \vspace{-8mm}
\end{figure*}

In this work, we introduce creases -- localized surface features that spontaneously nucleate via an elastic instability~\cite{trujillo2008creasing,chen2014controlled, dervaux2012mechanical,ciarletta2018matched} -- as rewritable motifs to gate, program and operate on droplets moving on soft interfaces. Remarkably, creases modify the elasto-capillary energy landscape giving rise to a long-range, repulsive force that halts droplets without physical contact. As a result, creases function as a passive yet reversible gate -- selectively allowing passage based on droplet size, as illustrated in Fig.~\ref{fig1}\textbf{a}. Combining experiments with theory, we show that the interaction force exhibits two distinct regimes, determined entirely by the the singular crease morphology. Using simple table-top fabrications, we demonstrate that such switchable crease gates can replicate a range of functional analogues to electrical circuits, including adaptive gating, path projection, hysteretic responses, memory register, signal modulation and digital logic.

\vspace{-5mm}
\section{Results}\vspace{-3mm}
\subsection*{Threshold dynamics}\vspace{-3mm}
\noindent We employ a soft-on-stiff bilayer platform, where a compliant gel layer is cured inside a pre-stretched, stiff polymer well and gets compressed upon release of the pre-stretch -- a common strategy for generating surface creases \cite{trujillo2008creasing,cai2012creasing,chen2014controlled}. To circumvent the uncontrolled nucleation of creases across the gel surface, we introduce a rectangular step ridge in the stiff base to focus compressive strain. This design reliably nucleate a single, linear crease directly above the ridge, as illustrated in the cross-sectional schematic in the middle panel of Fig.~\ref{fig1}\textbf{a} (also cf. Fig.~\blue{S1}). The soft layer consists of a silicone gel (Dow Corning CY52-276, A:B=1:1.3) with $E_{\mathrm{gel}}\simeq 3$ kPa, supported by a Vinyl Polysiloxane (VPS, Zhermack Elite Double) base  with $E_{\mathrm{base}}\simeq 1$ MPa (see Methods for further details). With the localized crease spanning the width of the gel surface, we deposit millimetric glycerin droplets onto the gel, driven toward the crease under gravity. Far from the crease, droplets move at a steady velocity, but as they approach the crease they decelerate. Near the crease, droplets' response bifurcates in a size dependent manner: smaller droplets stop at a distance $z=d_{\textup{stop}}$ upstream (cf. Fig.~\ref{fig1}\textbf{b}), while slightly larger droplets cross the crease (cf. Fig.~\ref{fig1}\textbf{c} and Supplementary Video \blue{1}). We observe that the halted droplets drift slowly parallel to the crease line (see Supplementary Video \blue{2} and Fig. \blue{S2}), indicating that they are not pinned to any surface defect but are instead halted by a long-range repulsion. 

To quantify the interaction dynamics, we track the position of each droplet’s leading edge relative to the crease, $z(t)$. The resulting trajectories in Fig.~\ref{fig1}\textbf{d} show that all droplets decelerate markedly as they approach the crease, deviating from a far-field steady velocity. For smaller droplets, the time to reach the crease increases until they are halted upstream without making contact (cf. Fig. \blue{S3} for magnified trajectories near the crease). Slightly larger droplets follow a more intricate sequence: the leading edge first touches down and pins, the bulk of the droplet rotates over, and finally the trailing edge depins and recovers the steady motion downstream. This sequence appears as two distinct kinks in the trajectory, highlighted on the orange curves in Fig.~\ref{fig1}\textbf{d}, and gradually vanish for large droplets where the crease has little effect on the dynamics. Across the full spectrum of drop sizes, these three behaviors are shown in the schematics of Fig.~\ref{fig1}\textbf{e}(i), and quantified in terms of the stopping distance $d_{\textup{stop}}$ in Fig.~\ref{fig1}\textbf{e}(ii). $d_{\textup{stop}}$ decreases monotonically with increasing 
$R$, vanishing at a critical radius $R_c\approx 0.8$ mm, above which droplets exhibit clean pass-through. Just below this threshold lies a narrow 0.05-mm band where the droplets pin to the crease with prolonged detachment. Because this pinning band is narrow, our discussion focuses on the dynamical switching between complete stoppage and clean passage. Further details of touchdown, detachment, and pinning band are provided in Supplementary Section \blue{2} and Fig.~\blue{S5}.

The threshold behavior demonstrates that the crease acts as an repulsive energy barrier, unlike conventional capillary defects that simply immobilize droplets upon contact. Instead, the crease is sensed from a distance, causing deceleration well before touching. This long-range influence suggests the role of the crease as a substrate-native, emergent defect that is switchable: releasing the compressive strain erases the fold, removes the barrier and restores uninterrupted transport. 

\begin{figure}[t]
    \begin{center}
    \centering
    \includegraphics[width=\columnwidth]{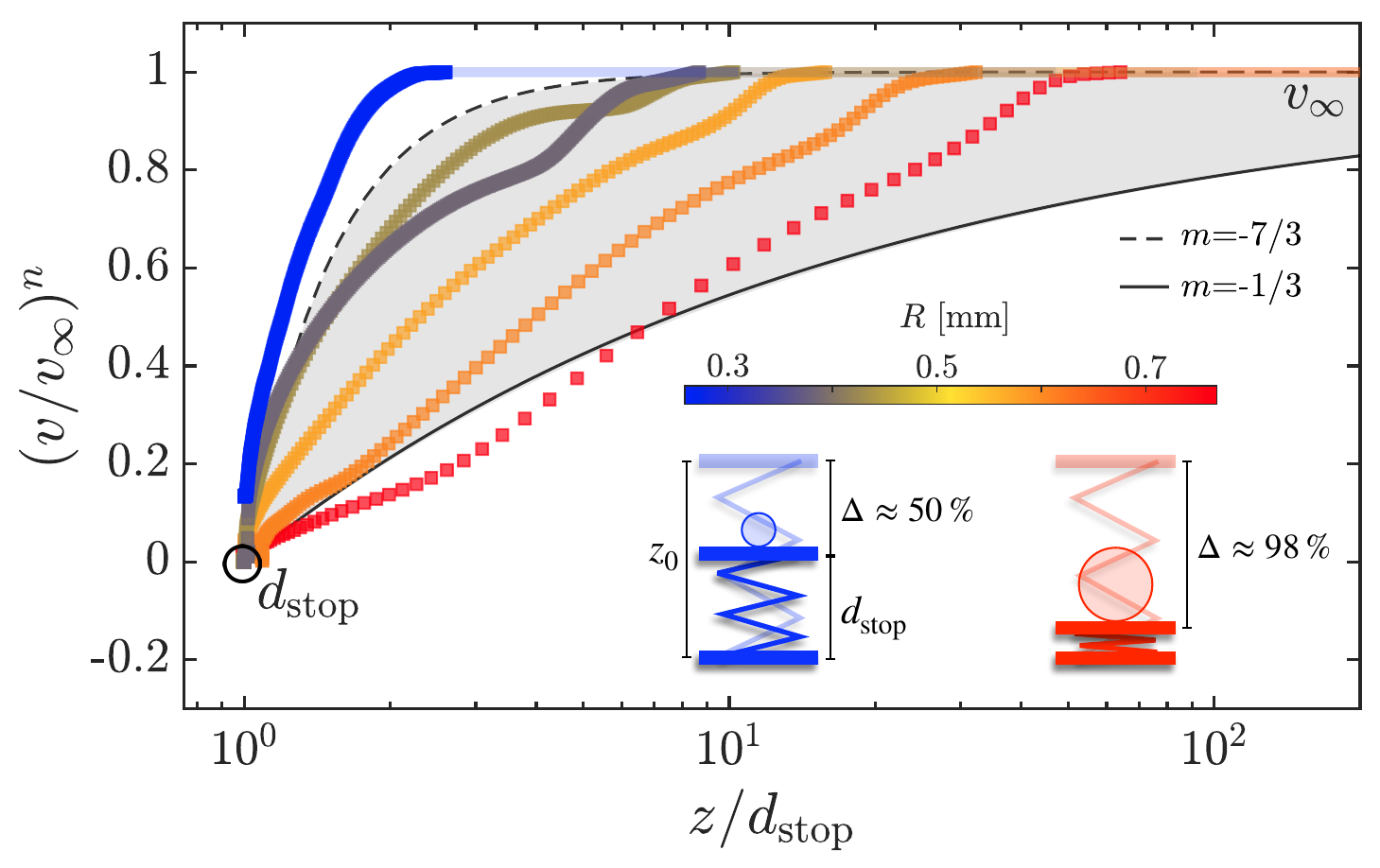}
    \caption{\justifying\textbf{Velocity profile of droplets arrested by a crease.} Droplet position $z$ is rescaled by the stopping distance $d_{\textup{stop}}$, and velocity $v$ is normalized by the far-field sliding speed $v_{\infty}$. The rheological exponent of the elastomer, $n$, is measured to be at 0.47. The rate at which droplets decelerate varies with size, as shown by the two dashed lines indicating different exponents for small and large droplets. The left inset illustrates the braking strain ($\Delta=\frac{z_0-d_{\mathrm{stop}}}{z_0}$) difference between the blue (small) vs red (large) drop for the elastocapillary spring. $z_0$ is the position where the drop starts to sense the crease, an influence cutoff point.}
    \label{fig2}
    \end{center}
    \vspace{-8mm}
\end{figure}

\begin{figure*}[h!tbp]
    \begin{center}
    \centering
    \includegraphics[width=0.9\textwidth]{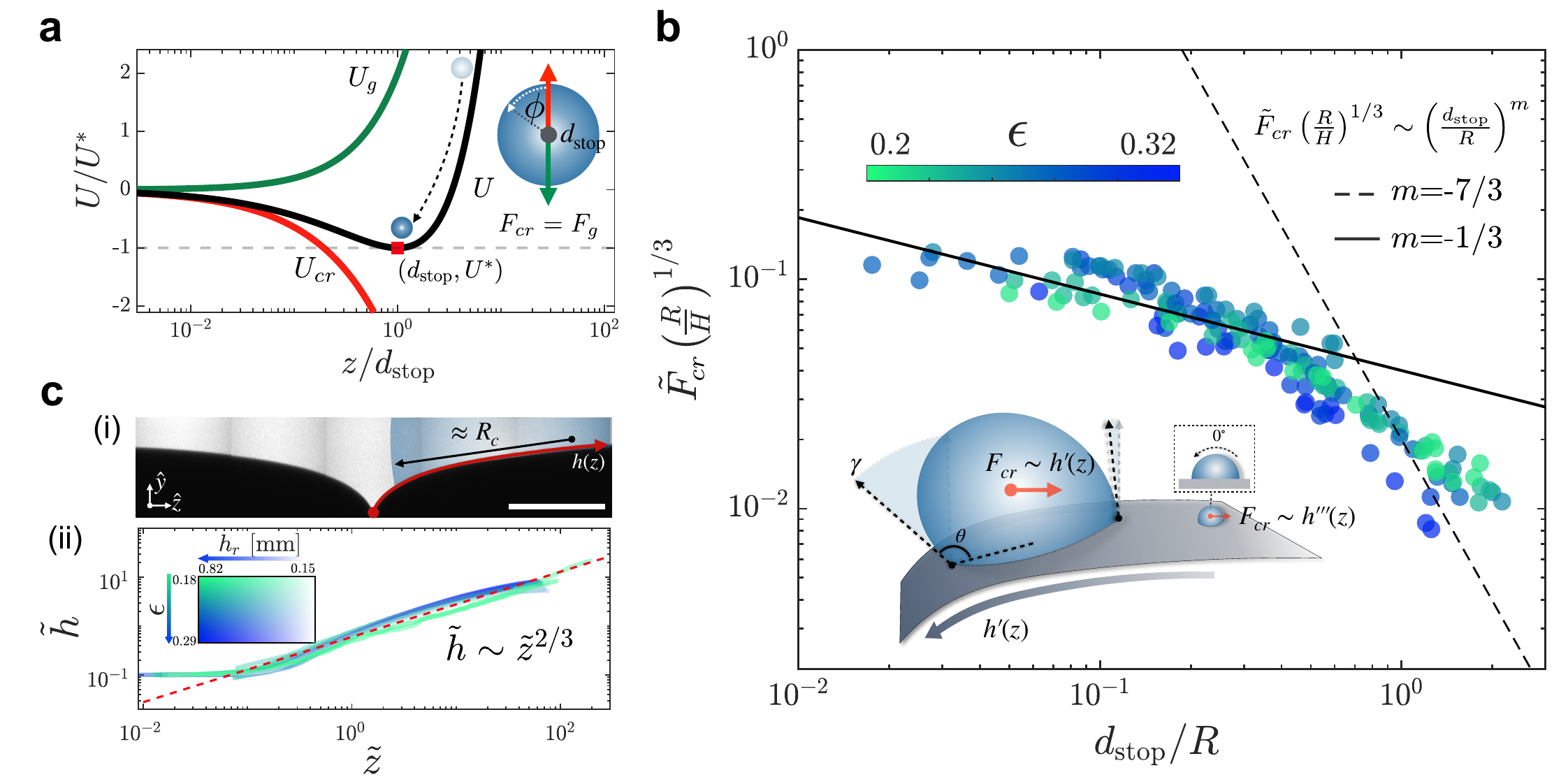}
    \caption{\justifying\textbf{Origin of crease repulsion.} \textbf{a.} Potential energy landscape for the drop-crease interactions. Inset schematic shows the instance of static equilibrium with the azimuthal angle definition. The green curve represents driving energy due to gravity, while the red curve represents the repulsive potential of the crease. A combination gives the potential well where the droplet stops. \textbf{b.} Normalized repulsive force vs. $d_{\mathrm{stop}}/R$, estimated from drop volume at $d_{\mathrm{stop}}$, for different compressive strains. Inset schematic shows the limiting cases of droplets at extreme sizes. The large drop experiences the crease deformation with significant rotation mismatch front vs. back, while the small drop feels minimal rotation. The solid line represents $m=-1/3$ and dashed line represents $m=-7/3$. Color bar denotes strain value. \textbf{c.} (i) Crease cross section image reconstructed from confocal microscopy. The scale bar here is 400 $\mu$m. Drop of radius $R_c$ is drawn on top for scale comparison. (ii) Normalized crease surface profiles $\bar{h}(\bar{z})$ as compressive strain $\epsilon$ varying from 0.18 to 0.29 and underlying step pattern ridge height $h_r$ varying from 0.15 to 0.82 mm.}
    \label{potential}
    \end{center}
    \vspace{-8mm}
\end{figure*}

\subsection*{Elastocapillary repulsion by creases}\vspace{-3mm}
\noindent In our experiments, droplets move slowly, with steady velocities ranging from 100 nm/s to 10 $\mu$m/s, typical for droplets on ultrasoft elastomers~\cite{CGS96,AS2020,Shanahan1994}. Inertia thus is negligible, and the steady motion arises from a balance between gravity and viscoelastic drag. The latter originates from dissipation localized in the substrate’s deformation at the contact line, known as the wetting ridge~\cite{LALLang96, karpitschka2015droplets}. The drag force takes a sublinear form $F_d = c v^n$, with exponent $n=0.47$ determined from rheological characterization (cf. Methods, SI Section~\blue{1,3}, and Fig.~\blue{S4,6}), and dominates over viscous losses within the droplet itself. As the droplet approaches a crease, it experiences an additional repulsive force akin to a nonlinear spring, $F_{cr} = k z^m$, where $z$ is the distance to the crease and $m<0$ ensures that the repulsion vanishes far away. With negligible inertia, the droplet dynamics reduces to an overdamped particle in the nonlinear force field, yielding:
\begin{equation}
  kz^{m}+cv^{n}=Mg. 
  \label{od eq}
\end{equation} Far from the crease ($z\rightarrow\infty$), the droplet's velocity $v_\infty$ is set by the balance $c v_\infty^n = Mg$. At full arrest, when the droplet rests at a distance $d_{\mathrm{stop}}$, gravity is balanced by the repulsion: $k d_{\mathrm{stop}}^{m} = Mg$. Normalizing $z$ by $d_{\mathrm{stop}}$ and $v$ by $v_{\infty}$ yields a dimensionless form of eq.~\eqref{od eq}:
\begin{equation}
  (z/d_{\mathrm{stop}})^{m}+(v/v_\infty)^n=1. 
  \label{odeq_ND}
\end{equation} 
We test this relation by plotting the rescaled velocity-distance data for droplets with radii below $R_c$ (0.25-0.75 mm) in Fig.~\ref{fig2}. All curves show a nonlinear decay deviating from the far-field steady motion. Smaller droplets experience a steeper decay, indicating that they engage a more nonlinear elastocapillary `spring' and stop sooner. In contrast, larger droplets decelerate more gradually, traversing much greater relative distances before stopping, as shown by the inset spring-compression schematics. The variation across sizes reflects differences in the effective spring exponent $m$. Indeed, the dashed and solid lines in Fig.~\ref{fig2} representing eq.~\eqref{od eq} for two exponents $m=-7/3$ and $m=-1/3$, roughly bracket the observed velocity profiles. To interpret these exponents and connect the dynamics to the underlying force law, we now turn to the static limit, where droplet motion ceases.

At static equilibrium, the dynamics reduce to an effective potential, $U(z)=Mgz-kz^{m+1}/(m+1)$, plotted in black in Fig.~\ref{potential}\textbf{a}, as the sum of gravity (green) and crease repulsion (red). Their interplay yields a shallow minimum at $[d_{\textup{stop}},\ U^*]=[(Mg/k)^{1/m},\ \frac{-m}{m+1}Mg d_{\mathrm{stop}}]$. This allows us to probe force–distance curve simply via droplet volume variation and obtained Fig.~\ref{potential}\textbf{b}. This reveals two distinct regimes depending on the normalized stopping distance $d_{\textup{stop}}/R$, far field region ($d_{\textup{stop}}/R\gg 1$) displays much steeper decay as compared to the near field ($d_{\textup{stop}}/R\ll 1$). $\tilde{F}_{cr}=F_{cr}/\gamma R$ is the non-dimensional force. $H=$ 3 mm is the constant gel thickness. The geometric factor of $\left(\frac{R}{H}\right)^{1/3}$ would be explained in the following section. Interestingly, the measured force remains unchanged across a range of compressive strains, though such strains are expected to alter the amplitude
of the crease deformation. In the next section, we derive the functional form of the spring constant $k$ and the nonlinear exponent $m$, and establish the scaling laws that explain the force-distance data of Fig.~\ref{potential}\textbf{b}.

\begin{figure*}[t]
    \begin{center}
    \centering
  \vspace{-10pt}
    \includegraphics[width=1.0\textwidth]{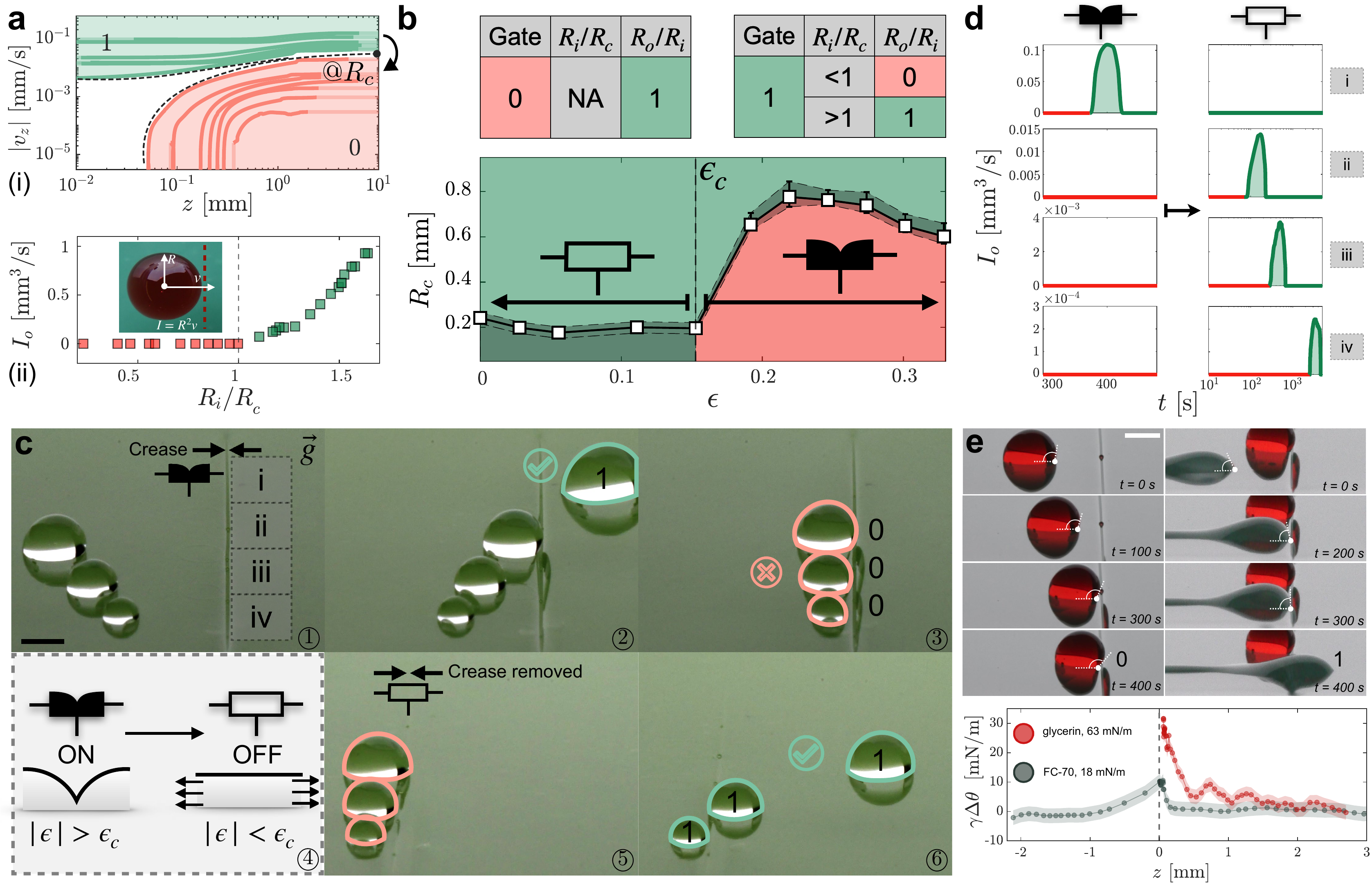}
    \caption{\justifying\textbf{Switchable drop filtration.} \textbf{a.} (i) Zoomed view on the drop dynamics bifurcations in the $v-z$ parameter space. (ii) Output current $I_o$ as a function of $R_i/R_c$. Inset shows current definition as $I=R^2v$. \textbf{b.} Upper panel shows the drop signal truth table. Lower panel shows the critical drop size $R_c$ as a function of strain. Red regime is associated with state 0 for signal; green regime is associated with state 1. \textbf{c.} Demonstration of the crease as an autonomous drop signal filter, with switchability. The temporal sequence is marked by \textcircled{1} to \textcircled{6}. Four signal readout locations are labeled from i to iv. See Supplementary Video \blue{3}. \textbf{d.} Signal readout via image analysis (output current $I_o$) for the four channels in \textbf{c} as a function of time. \textbf{e.} Gating based on surface tension. Glycerin drop in red is halted; FC-70 drop in grey of the same volume passes through. Upper panels show the experimental sequences (also see Supplementary Video \blue{4}), and the lower plots show the calculation of effective force per length felt from the leading edge rotation $\gamma\Delta\theta$. All scale bars here are at 1 mm.} 
    \label{filter}
    \end{center}
    \vspace{-8mm}
\end{figure*}

\vspace{-3mm}
\subsection{Force-distance relation}\vspace{-3mm} 
\noindent The force–distance measurements above show that droplets can be halted at distances up to three times their radii. This long-range sensing is mediated by the elastic energy of the soft substrate. Both the crease and the droplet generate extended deformations over hundreds of microns: the crease forms a stationary line indentation, while the droplet carries a mobile wetting ridge. We capture this interaction within an energy framework, treating the droplet as a perturbation to the creased substrate. In this formulation, the total energy of the coupled system is expressed as
\begin{equation}
E=\mathcal{E}_{ec}[h]+\int T(\mathbf{x}-\mathbf{x}_d)\, h(\mathbf{x})\mathrm{d}\mathbf{x}.
\label{S7}
\end{equation} Here $\mathcal{E}_{ec}$ is the elasto–capillary energy of the interface $h(\mathbf{x})$ with a pre-existing crease at $z=0$, and the second term accounts for the work of the droplet's traction $T(\mathbf{x})$ centered at $\mathbf{x}_d$. The repulsive force on the droplet follows from differentiating the energy with respect to the relative separation of crease and droplet, $\mathbf{F}=-\partial E/\partial{\mathbf{x}_d}$~\cite{pandey2018hydrogel}. Crucially, only the third term on the right in eq.~\eqref{S7} which couples droplet traction to the interface shape contributes to this force (cf. SI Section~\blue{4} for details).\\
\noindent\underline{\textit{Far-field scaling}} - The capillary traction $T(\mathbf{x})$ depends on the droplet footprint and contact angle, both of which evolve near a crease and are therefore unknown \textit{a priori}. In the far-field limit, however, the droplet retains the configuration of one on a flat substrate, yielding as axisymmetric traction in a polar coordinates $(\rho, \phi)$ centered at the droplet,
$T(\rho, \phi)=\gamma\sin\theta\left[\delta(\rho-R)-R^{-1}\Theta(R-\rho)\right]$. Here $\theta$ is liquid contact angle, $\delta(\rho)$ represents the upward line force at the contact line, and $\Theta(\rho)$ the downward Laplace pressure distributed over the footprint. Because the interface profile varies slowly at distances $d_{\textup{stop}}\gg R$, we expand $h$ about the droplet center. Symmetry of the traction field removes the monopole and dipole terms, leaving the quadrupole as the leading contribution (see SI Section~\blue{4} for detailed calculation). It follows that the droplet couples to the gradient of curvature of the crease profile, giving a repulsion,
\begin{equation}
    F_{cr} \sim-\gamma\sin\theta R^3 h'''(d_{\textup{stop}}).
    \label{eq4}
\end{equation} Direct measurement of interface shape around the crease using confocal microscopy enables an explicit force-distance scaling following eq.~\eqref{eq4} (see Methods for details). Fig.~\ref{potential}\textbf{c}(i) shows a representative confocal slice of the crease, while \ref{potential}\textbf{c}(ii) compiles rescaled crease profiles obtained for a range of compressive strains ($\epsilon$) and mold step heights ($h_r$). It shows that despite variations in amplitude, the near‑tip morphology is self‑similar: using $\tilde{z}=zl^{-3/4}A_c^{-1/4}$ and $\tilde{h}=h(lA_c)^{-1/2}$ (with $l=\gamma_{\mathrm{gel}}/G$ being the substrate's elastocapillary length, and $A_c$ is the crease amplitude measured at the tip), all profiles collapse onto a cusp shape $\tilde{h}\sim \tilde{z}^{2/3}$~\cite{karpitschka2017cusp, van2021pinning}. Here $A_c$ connects the local morphology to the excess compressive strain $\Delta\epsilon=\epsilon-\epsilon_c$ and gel thickness $H$ through $A_c\sim H(\Delta\epsilon)^{1/2}$~\cite{essink2023crease}. The intermediate 2/3 cusp regime is cut off at small $\tilde{z}$ by elastocapillary regularization near the tip and flattens at large $\tilde{z}$ by finite-size effects. Plugging the crease shape in eq.~\ref{eq4} we find the far‑field force–distance scaling
\begin{equation}
\tilde{F}_{cr} \sim \left(\tfrac{R}{H}\right)^{-1/3} (\Delta\epsilon)^{1/6}\, \tilde{d}_{\textup{stop}}^{-7/3},
\label{far-field force scaling}
\end{equation} where $\tilde{d}_{\textup{stop}}=d_{\textup{stop}}/R$ is the dimensionless distance. Eq.~\eqref{far-field force scaling} is plotted as a dashed line in Fig.~\ref{potential}\textbf{b} and rationalizes the steep decay in the force data for smaller droplets.\\
\noindent\underline{\textit{Near-field scaling}} - In the near field, we observe that halted droplets tilt forward and their contact angles vary across the footprint, as shown in the schematics in Fig.~\ref{potential}\textbf{b} inset. This tilt reflects a rotation of the wetting ridge~\cite{karpitschka2016liquid,karpitschka2015droplets}: the local slope of the crease $h'$ projects onto the outward normal of the contact line, tipping the wetting ridge forward. The resulting surface tension imbalance along the footprint leaves a residual force in $\hat{z}$, whose projection onto the axis normal to the crease yields a net repulsion. This geometric coupling produces a $\cos^2\phi$ dependence in the force integral, yielding
\begin{equation}
F_{cr}=\gamma\sin\theta\, R\int_0^{\pi}h'(z)\cos^2\phi\,\mathrm{d}\phi.
\label{near-field force}
\end{equation} Here $\phi$ is the azimuthal angle defining the droplet's circular footprint (cf. Fig.~\ref{potential}\textbf{a}), and $z=d_{\textup{stop}}+R(1+\cos\phi)$ locates a point on the contact line relative to the crease (See Supplementary Fig.~\blue{S7} for schematics and Section \blue{4} for calculation details). Plugging in the crease shape of the crease in eq.~\eqref{near-field force}, we find the near-field force-distance scaling
\begin{equation}
    \tilde{F}_{cr}\sim \left(\tfrac{R}{H}\right)^{-1/3}(\Delta\epsilon)^{1/6}\, \tilde{d}_{\textup{stop}}^{-1/3}.
\label{near-field force scaling}
\end{equation} This result is plotted as a solid line in Fig.~\ref{potential}\textbf{b}, closely matches the data and explains the velocity decay of larger droplets in Fig.~\ref{fig2}. Because the prefactor of eqs.~\eqref{far-field force scaling} and~\eqref{near-field force scaling} depend only weakly on the excess strain $\Delta\epsilon$, data obtained at different compression levels collapse onto a single master curve. 

\begin{figure}[t]
    \begin{center}
    \centering
    \includegraphics[width=\columnwidth]
    {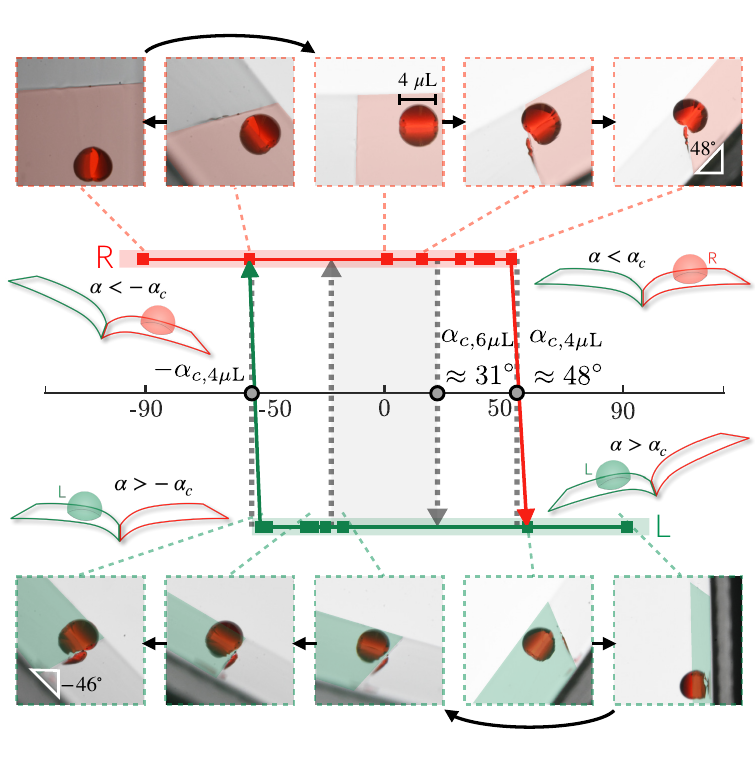}
    \vspace{-8mm}
    \caption{\justifying\textbf{Angular hysteron.} Experimental sequences and the resulting I/O map for the hysteron is shown for a 4 $\mu L$ drop. The top panel experiments demonstrate the drop residence on the right ($\mathrm{R}$) side; the bottom panels demonstrate the opposite residence on the left ($\mathrm{L}$) side. The transition occurs at the $\pm\alpha_c\approx\pm 48^{\circ}$ (shown in the experimental frames). Inner loop shows a contraction to $\pm\alpha_c\approx\pm 31^{\circ}$ via drop volume tuning. See this I/O map in Fig. \blue{S}9.}
    \label{Hysteron}
    \end{center}
    \vspace{-8mm}
\end{figure}

\vspace{-5mm}
\subsection{Crease as switchable drop filter}\vspace{-3mm}
\noindent Having established the interaction laws, we now show how to harness such nonlinearities to achieve functional control over drop traffic. The nonlinear potential landscape bifurcates droplet trajectories near $R_c$, naturally creating an autonomous size-based filter without external feedback; the bifurcation point in the $v-z$ parameter space is shown in Fig.~\ref{filter}\textbf{a}(i). We define the drop current as a simplified volumetric flow rate of $I=R^2v$. With $R_o$ and $R_i$ denoting the output and input drop radius respectively, the output current $I_o$ is shown in Fig.~\ref{filter}\textbf{a}(ii), demonstrating $I_o$ nonlinearly rising with $R_i-R_c$ above the cutoff size $R_c$. 

The crease gate is activated only above the critical strain $\epsilon_c\approx$ 0.18. The gate can therefore be toggled between an OFF (no crease) and ON (crease) state. The input-output (I/O) maps for both configurations are shown in the top-panel of Fig.~\ref{filter}\textbf{b}. The lower panel of \ref{filter}\textbf{b} shows the variation of the critical radius $R_c$ as a function of applied strain. Below $\epsilon_c$ (OFF state), the output $R_o/R_i$ is close to an unconditional 1, and the signal is fully transmitted downstream. At $\epsilon=\epsilon_c$, $R_c$ jumps sharply (ON state), reflecting the sudden emergence of the elastic barrier, and subsequently reduces slightly over the strain range we explored. 

Thus a key advantage of a crease gate is its reversibility: releasing the compression erases the crease and enable reconfiguration, see Fig.~\ref{filter}\textbf{c} and Supplementary Video \blue{3}. In the ON state (\textcircled{1}-\textcircled{3}), channels ii--iv are blocked; after de-compression, the gate turns off, and previously blocked channels transmit fully $R_o/R_i$=1 (\textcircled{4}-\textcircled{6}). The corresponding current outputs are summarized in Fig.~\ref{filter}\textbf{d}. Small residual scars may remain on the surface after reversal\cite{van2021pinning, liu2019elastocapillary}, producing uncertain $R_c$ values below 0.2 mm (gray regions in Fig.~\ref{filter}\textbf{b}). Functionally, the crease thus acts as an internal activation barrier, analogous to a synapse \cite{wilson1972excitatory} or p-n junction \cite{shockley1949theory}. The strain-controlled switching provides an external knob, as reflected in the sigmoid-like variation of $R_c$ with compression in Fig.~\ref{filter}\textbf{b}. This parallels field-effect transistor (FET), where current flows to the drain only as the gate-to-source voltage $V_{GS}$ exceeds the threshold $V_{th}$.

\begin{figure*}[h!tbp]
    \begin{center}
    \centering
  \vspace{-10pt}
    \includegraphics[width=1\textwidth]{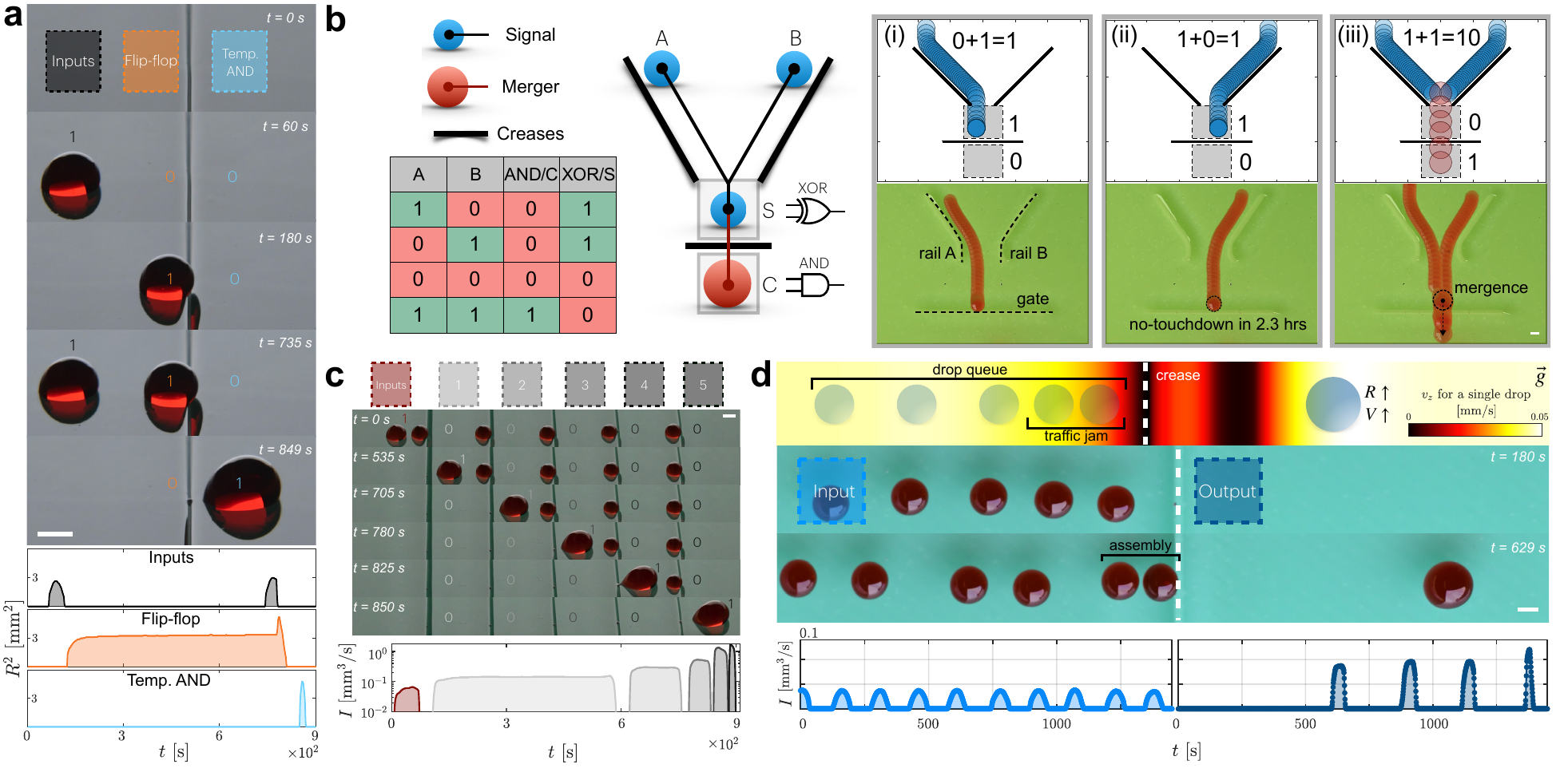}
    \caption{\justifying\textbf{Drop signal operations.} \textbf{a.} Experimental sequence for sequential logic realization of flip-flop (upstream) and temporal AND (downstream) units. The current responses at the input, flip-flop, and AND locations are plotted in the lower panels. \textbf{b.} Truth table (left), gate design (center), simulated sequence (upper panels of (i)-(iii)) and experiments (lower panels of (i)-(iii)) for a half adder, constructed from AND and XOR gates designated to be the immediate upstream and downstream of the end crease. Inputs are guided by crease rails A and B. See full sequence in Fig.~\blue{S10}. \textbf{c.} Signal cascade demonstrated from five crease gating units connected in series. Droplets are halted above each and flipped by one additional input signal upstream, which triggers the avalanche responses with amplifying currents downstream. \textbf{d.} Signal pulse modulation with different drop current pulses (varying in frequency and amplitude) upstream and downstream of the crease. Upper panel is the velocity map along $z$, which shows the origin of traffic jamming for the drop queue to cross the gate. Middle panel is the experiments showing self assembly upstream, and the lower panel shows the current responses. All scale bars are at 1 mm. See all the above videos in Supplementary Video \blue{6}.}
    \label{logic}
    \end{center}
    \vspace{-8mm}
\end{figure*}

The crease gate also displays multi-modality, operable with variables beyond sizes. Since the interaction arises from elastocapillarity, surface tension can also be used to tune the critical cutoff, enabling $\gamma$ gating module. We demonstrated this in Fig.~\ref{filter}\textbf{e} and Supplementary Video \blue{4}, showing two drops of the same volume, glycerin in red with $\gamma_{\mathrm{gl.}}\approx$ 64 mN/m vs. FC-70 (3M Fluorinert Electronic Liquid) in gray with $\gamma_{\mathrm{FC}}\approx$ 18 mN/m. For these two cases, the leading-edge apparent angle rotation $\Delta \theta$ during the approach are tracked. From this rotation, we approximate the tension (force per length) at the leading edge felt as $\gamma h'_{\mathrm{max}}\sim \gamma\Delta \theta$, since $h'_{\mathrm{max}}\approx \Delta\theta$ under small rotations. The result is shown in the lower panel, explaining the pass-through $R_o/R_i=$1 for FC-70 and stoppage $R_o/R_i=$0 for glycerin. The gating size $R_c$ thus decreases with $\gamma$.  

\vspace{-5mm}
\subsection{Angular hysteron}\vspace{-3mm} 
\noindent Thus far, we have established that localized creases act as a switchable automatic drop filters with transistor-like functions. In all previous cases, the surface was held parallel to gravity, with a single governing knob of size cutoff 
$R_c$. Here we further design the bistability by playing with a tilted substrate of angle $\alpha$ (with $\alpha=90^{\circ}$ corresponding to the gravity-parallel case). The droplet's residence on either side of the crease (left ($\mathrm{L}$) and right ($\mathrm{R}$)) is togglable by $\alpha$, as demonstrated in the I/O response curves of Fig.~\ref{Hysteron}, accompanied by experimental images and schematics. Above a critical tilt $\alpha_c\simeq 48^{\circ}$, the droplet spontaneously switches from $\mathrm{R}$ to $\mathrm{L}$, while the reverse transition occurs only at the opposite tilt $-\alpha_c$, forming a hysteresis loop. We refer to this bistable element, defined by a return-point memory of angle, as an angular hysteron. The memory point $\alpha_c$ is controllable by drop size, thereby stretching or contracting the hysteresis boundary. Supplementary Section \blue{7} and Fig. \blue{S9} demonstrate this tunability: increasing the droplet volume from 4 $\mu$L to 6 $\mu$L contracts the loop, with $\alpha_c$ decreasing from 48$^{\circ}$ to 31$^{\circ}$, expected from $\mathrm{sin}(\alpha_{c,\ 4 \mu \mathrm{L}})/\mathrm{sin}(\alpha_{c,\ 6 \mu \mathrm{L}})\approx 6/4$. This dynamics is identical to that of a relay hysteron in the Preisach model \cite{semenov2024preisach}, displaying history dependence that is the basis of data encoding and storage applications, from the earlier single-dial combination lock \cite{lindeman2025generalizing} to modern ferroelectric RAM devices \cite{semenov2024preisach}. 

\vspace{-5mm}
\subsection{Drop signal operations}\vspace{-3mm} 
\noindent The angular hysteron captures a fundamental notion of history-dependence for memory register. Now we extend the framework to demonstrate logical signal operations by embedding state responses within structured temporal or spatial schemes. The quantized nature of drops allows them to serve as discrete pulse inputs. Within a single drop channel towards the gate, we can exploit droplet coalescence. In this regard, we designate readout patches at specific positions along $z$ (cf. Fig.~\ref{logic}\textbf{a}). In the patch immediately before the gate, each incoming pulse toggles the stored state between 0 and 1, thereby realizing the truth table of a toggle flip-flop. The state persists until another droplet arrives, realizing a different variant of memory. In contrast, the downstream patch simply registers passage, and thus outputs only when two droplets arrive sequentially -- functionally equivalent to a temporal AND unit. The measured current responses at the bottom of Fig.~\ref{logic}\textbf{a} confirm both operations.

To move beyond sequential operations based on historical states, we designed two separate input channels above a single crease to realize combinational logic operations. The paths of subcritical droplet bits were controlled using additional creases as guiding rails, since droplet bits below cutoff cannot cross them. These patterned creases effectively project the incoming droplets onto the terminal crease. Please see further demonstrations and characterizations of crease path guidance in Supplementary Section \blue{5}, Fig. \blue{S8} and Supplementary Video \blue{5}, which shows droplets travel along the crease with their distinct $d_{\mathrm{stop}}$ spacing, proving again droplets are not pinned but instead guided on a potential well channel. 

We first utilized numerics to design a half-adder setup. The simulation solves for the droplets' speeds and locations from the ODE with gravity and crease repulsion. For details, see Supplementary Section \blue{6}. The resulting design, shown in left panels of Fig.~\ref{logic}\textbf{b}, demonstrates how spatially separated inputs transform the terminal crease into a natural rendition of a half adder unit. The upstream and downstream locations are registered as AND and XOR gate respectively. The XOR readout patch produces a signal 1 only for odd inputs of 1, and serves as the Sum (S) in signal readout, while the AND patch produces 1 for even input of 1 and perform the Carry (C). The panels (i)--(iii) in \ref{logic}\textbf{b} show composite images from both the numerics and experiments demonstrating the 3 nontrivial cases of 0+1=1, 1+0=1, and 1+1=10. Additional types of operation (NAND, full adder) are designed via simulation, discussed and shown in Supplementary Section \blue{7} and Fig. \blue{S11}. 

The above demonstrations realize both categories of Boolean logic: sequential and combinational. With further engineering, we can potentially split droplets to preserve signals and achieve scalable logic. Drop queue across the gate also carries two distinct features beyond logic: built-in gain and event interval compression, since merger naturally amplifies both speed and size as $I=R^2v\sim R^{6.25}$ with $n=$ 0.47. The power law of $v$ with size here is specific to the current material and geometry, and is subjected to change (see validation in Supplementary Section \blue{1} and Fig. \blue{S4}). We now demonstrate below two systems emerging from this property of innate gain. 

The first is a cascading amplifier, realized by connecting multiple crease units in series as shown in Fig.~\ref{logic}\textbf{c}. Five drops are first halted upstream by the five creases, and an additional droplet is introduced to trigger a series of downstream toggles that generate an avalanche of current spikes, signature of hysteretic cascade. Crossing the $k_{th}$ crease scales the input drop radius by a factor of $(k+1)^{1/3}$; therefore, from the above calculation of $I\sim R^{6.25}$, $I_k/I\approx (k+1)^{2}$. For $k=5$, we have a roughly 36 fold gain by scaling, slightly above the experimental measure of 24. The events intervals downstream are also compressed exponentially after each subsequent gate. Overall, the run-away sequence in \ref{logic}\textbf{c} constitutes a self-reinforcing hysteron chain with superlinear current gain, reminiscent of magnifying domino and photonic cascading amplifiers \cite{van2013domino,bilal2017bistable,van2010domino,mcarthur2023demonstration}. 

Building upon this, we can further modify the dynamic properties of a drop queue (a periodic pulse signal) to achieve pulse modulation, as shown in Fig.~\ref{logic}\textbf{d}. Owing to the long-range interaction, the incoming queue encounters a `traffic jam' near the gate, as visualized in the velocity map for droplets $R\approx0.8$ mm in Fig.~\ref{logic}\textbf{d} top panel. Two distinct traffic bands emerge: the first is associated with the repulsion, while the second is with detachment. Experiments in \ref{logic}\textbf{d} middle panel reveal that in the repulsion band, droplets self-assemble upstream, and the crease subsequently releases a boosted pulse at higher speed. This reshapes the input signal into a fully modulated waveform (Fig.~\ref{logic}\textbf{d}, bottom panel). From the above scaling derivations, pulse amplitude ($A$), frequency ($f$), and pulse width ($T$) are modified as $A_o/A_i\sim j^2$, $f_o/f_i\sim 1/j$, and $T_o/T_i\sim j^{-1.4}$, where $j$ is the number of mergers initiated. Thus, crease gates thus boost droplet momentum in discrete notches, constituting a drop waveform modulator (see Supplementary Video \blue{6}).

\vspace{-3mm}
\section{Discussion}\vspace{-3mm}
\noindent Our findings reveal that creases -- ubiquitous elastic instabilities at soft interfaces -- act as long-range elastocapillary barriers for droplets. This establishes a previously unrecognized class of interactions in which engineered singular, elastic morphologies repel droplets at distances several times their radii. The droplet response shows a sharp threshold: below a critical radius droplets halt upstream, while above it they pass through. Importantly, this threshold is tuned by the onset of creasing, so that small changes in compressive strain around the instability produce large modulation in $R_c$. The crease thus serves as a nonlinear control knob, directly coupling substrate mechanics to droplet dynamics and linking singular geometry to selective transport.

Beyond this fundamental advance, the repulsive force is sufficiently strong, tunable, and reversible to be exploited as a functional unit for discrete-phase fluidics. By toggling creases through strain, we achieve switchable gates that filter droplets by size or surface tension, demonstrate angular hysterons with memory, and realize capsule routing, pulse modulation, signal amplification, and Boolean logic operations. These operations bypass some of the inherent limitations of traditional digital microfluidics (DMF), which often demand complex actuation, intensive resources, costly fabrication, and cannot adapt to unplanned experiments once routes are predetermined \cite{wang2020electrowetting, faustino2016biomedical, paratore2022reconfigurable}. In contrast, crease gates are reconfigurable and potentially rewritable via strain gradients, yet require only benchtop synthesis. Further integration and engineering can potentially feed droplets back into the gate to reprogram the crease state, split droplets and introduce clocking. Such capacity would allow droplets to both sense and modify the transport program, fully realizing cascadability, feedback, and synchronicity to achieve scalability \cite{katsikis2015synchronous, prakash2007microfluidic}. Overall, the crease gate opens up a novel, surface-native, low-entry-barrier route for interfacial information processing, enabling potential applications in reconfigurable microfluidics \cite{paratore2022reconfigurable}, interfacial logic \cite{abdelgawad2009digital}, condensations \cite{sokuler2010softer}, self assembly \cite{timonen2013switchable}, microfabrication \cite{srinivasarao2001three}, and other analytical chemistry automation \cite{stone2004engineering,prakash2008surface,tuteja2008robust,abdelgawad2009digital,seemann2011droplet,nosonovsky2009superhydrophobic}.

\section{Methods}

\subsection{Dual layer crease localizations}
\noindent The stiff polymer case is made of Vinyl
Polysiloxane (VPS) rubber (Zhermack Elite Double 22 or 32), and is first molded into shape from 3D-printed PLA molds. The design is mainly an 90 mm by 80 mm rectangular block, with a 60 mm by 50 mm well, that is 3 mm in depth. To localize the formation of a single crease, rectangular step ridges of various length or angle are added onto the bottom of the well. Cross sections of the ridges are around 0.6 by 0.6 mm. Creases would be nucleated upon locations of the ridges under compression due to local variation in thickness. After molding the VPS polymer, this polymer case is stationed onto two linear stages in parallel via customarily machined clamps, and stretched uniaxially to $\sim$150$\%$, allowing up to 30$\%$ compressive strain after. The full setup schematics can be found in Supplementary Fig. \blue{S1}\textbf{a}. After stretching out the base, the soft gel composition is made from 1:1.3 of Dowsil CY 52-276, A and B; it is then degassed in a vacuum dessicator for 5-10 minutes and deposited into the well in the base. The gel was cured at 75$^\circ$ for 5 hours, with additional 2 days of resting time afterward. Upon compression above a critical strain $\epsilon_c\simeq 18\%$, the step structure triggers nucleation of a single, well-aligned crease.  We conducted oscillatory rheology to extract mechanical properties of the gel (see Rheology section below). The biaxial stretch is created with a similar setup, shown in Fig. \blue{S1}\textbf{b}, to realize the half-adder experiment. Experimental images of the basic uniaxial and biaxial setups shown in Fig. \blue{S1}\textbf{c} and \textbf{d} respectively. Images are taken with Nikon Z5 cameras with extension lenses. We direct the cameras towards the sample from both the $\hat{y}$ (head-on) and $\hat{x}$ (side) directions. A time-lapse video is used to capture the drop motions, with typical frame rate of 0.1 to 1 frame per second.

\subsection{Rheology}
\noindent Rheology of the gel sample is conducted with DHR3 Rheometer from TA Instrument. An 8-mm-diameter flat rotation head are used for shearing. We applied a frequency sweep on samples with various A-B mixing ratios $r_{\mathrm{AB}}$ and extracted the $G'$ and $G''$ curves as shown in Fig. \blue{S6}, with fitting to calculate the parameters of $G_0$, viscoelasticity exponent $n$, and time scale $\tau$, using the Chasset-Thirion equation as detailed in the Supplementary Section \blue{3}. The shearing strain is varied from 1 -- 20$\%$, yielding similar values with minimal error bars. 

\subsection{Confocal microscopy}
\noindent For imaging, we placed the entire stage, with the gel, base layer, and the clamp under an upright Zeiss LSM880 confocal/multiphoton microscope. A special gel sample is used where the ridge varies height from 0.1 to 0.9 mm in the $\hat{x}$ direction, instead of being fixed at 0.6 mm. The gel layer thickness is kept at 3 mm. The laser line used are 488 excitation with emission 500-606 nm. We used a ophthalmologic solution of fluorescein (AK-Fluor 25$\%$ Akorn brand) at concentration of around 10 $\mu$M in Phosphate-buffered saline (PBS) to submerge the crease profile for visualization. By varying strain on the specially designed sample, we sweep the parameter space of strain $\epsilon$ and ridge height $h_r$ for the crease's cross-sectional profiles. 

\section{Acknowledgment}
\noindent We thank Rebecca M. Williams for her contributions in facilitating the confocal microscopy, and Krishan Badrie for his contributions in experiments. Confocal data was acquired through the Cornell Institute of Biotechnology's Imaging Facility, with NIH S10OD018516 funding for the shared Zeiss LSM880 confocal/multiphoton microscope. G.L. acknowledges funding from the Syracuse Office of Undergraduate Research \& Creative Engagement (The SOURCE). Z.W. and A.P. acknowledge start up funding from Syracuse University. S.K. acknowledges funding from the German Research Foundation (DFG, Project No. 422877263).

\section{Author contributions}
\noindent Z.W., S.K., A.P., conceived the idea. Z.W., G.L. designed and performed experiments. Z.W., A.P. and S.K. developed the dynamical model and A.P. and S.K. developed the theoretical model. Z.W. developed numerical calculations. All authors wrote the paper collectively. 

 \bibliographystyle{ieeetr}
 \bibliography{main}

\begin{thebibliography}{10}

\bibitem{fredkin1982conservative}
E.~Fredkin and T.~Toffoli, ``Conservative logic,'' {\em International Journal of theoretical physics}, vol.~21, no.~3, pp.~219--253, 1982.

\bibitem{preston2019digital}
D.~J. Preston, P.~Rothemund, H.~J. Jiang, M.~P. Nemitz, J.~Rawson, Z.~Suo, and G.~M. Whitesides, ``Digital logic for soft devices,'' {\em Proceedings of the National Academy of Sciences}, vol.~116, no.~16, pp.~7750--7759, 2019.

\bibitem{prakash2007microfluidic}
M.~Prakash and N.~Gershenfeld, ``Microfluidic bubble logic,'' {\em Science}, vol.~315, no.~5813, pp.~832--835, 2007.

\bibitem{sitti2021physical}
M.~Sitti, ``Physical intelligence as a new paradigm,'' {\em Extreme Mechanics Letters}, vol.~46, p.~101340, 2021.

\bibitem{faustino2016biomedical}
V.~Faustino, S.~O. Catarino, R.~Lima, and G.~Minas, ``Biomedical microfluidic devices by using low-cost fabrication techniques: A review,'' {\em Journal of biomechanics}, vol.~49, no.~11, pp.~2280--2292, 2016.

\bibitem{schutzius2015spontaneous}
T.~M. Schutzius, S.~Jung, T.~Maitra, G.~Graeber, M.~K{\"o}hme, and D.~Poulikakos, ``Spontaneous droplet trampolining on rigid superhydrophobic surfaces,'' {\em Nature}, vol.~527, no.~7576, pp.~82--85, 2015.

\bibitem{daniel2001fast}
S.~Daniel, M.~K. Chaudhury, and J.~C. Chen, ``Fast drop movements resulting from the phase change on a gradient surface,'' {\em Science}, vol.~291, no.~5504, pp.~633--636, 2001.

\bibitem{bjelobrk2016thermocapillary}
N.~Bjelobrk, H.-L. Girard, S.~Bengaluru~Subramanyam, H.-M. Kwon, D.~Qu{\'e}r{\'e}, and K.~K. Varanasi, ``Thermocapillary motion on lubricant-impregnated surfaces,'' {\em Physical Review Fluids}, vol.~1, no.~6, p.~063902, 2016.

\bibitem{chen2016continuous}
H.~Chen, P.~Zhang, L.~Zhang, H.~Liu, Y.~Jiang, D.~Zhang, Z.~Han, and L.~Jiang, ``Continuous directional water transport on the peristome surface of nepenthes alata,'' {\em Nature}, vol.~532, no.~7597, pp.~85--89, 2016.

\bibitem{kusumaatmaja2007controlling}
H.~Kusumaatmaja and J.~Yeomans, ``Controlling drop size and polydispersity using chemically patterned surfaces,'' {\em Langmuir}, vol.~23, no.~2, pp.~956--959, 2007.

\bibitem{abdelgawad2009digital}
M.~Abdelgawad and A.~R. Wheeler, ``The digital revolution: a new paradigm for microfluidics,'' {\em Advanced Materials}, vol.~21, no.~8, pp.~920--925, 2009.

\bibitem{abdelgawad2008all}
M.~Abdelgawad, S.~L. Freire, H.~Yang, and A.~R. Wheeler, ``All-terrain droplet actuation,'' {\em Lab on a Chip}, vol.~8, no.~5, pp.~672--677, 2008.

\bibitem{vialetto2017magnetic}
J.~Vialetto, M.~Hayakawa, N.~Kavokine, M.~Takinoue, S.~N. Varanakkottu, S.~Rudiuk, M.~Anyfantakis, M.~Morel, and D.~Baigl, ``Magnetic actuation of drops and liquid marbles using a deformable paramagnetic liquid substrate,'' {\em Angewandte Chemie International Edition}, vol.~56, no.~52, pp.~16565--16570, 2017.

\bibitem{timonen2013switchable}
J.~V. Timonen, M.~Latikka, L.~Leibler, R.~H. Ras, and O.~Ikkala, ``Switchable static and dynamic self-assembly of magnetic droplets on superhydrophobic surfaces,'' {\em Science}, vol.~341, no.~6143, pp.~253--257, 2013.

\bibitem{ichimura2000light}
K.~Ichimura, S.-K. Oh, and M.~Nakagawa, ``Light-driven motion of liquids on a photoresponsive surface,'' {\em Science}, vol.~288, no.~5471, pp.~1624--1626, 2000.

\bibitem{sun2019surface}
Q.~Sun, D.~Wang, Y.~Li, J.~Zhang, S.~Ye, J.~Cui, L.~Chen, Z.~Wang, H.-J. Butt, D.~Vollmer, {\em et~al.}, ``Surface charge printing for programmed droplet transport,'' {\em Nature materials}, vol.~18, no.~9, pp.~936--941, 2019.

\bibitem{pit2015high}
A.~M. Pit, R.~de~Ruiter, A.~Kumar, D.~Wijnperl{\'e}, M.~H. Duits, and F.~Mugele, ``High-throughput sorting of drops in microfluidic chips using electric capacitance,'' {\em Biomicrofluidics}, vol.~9, no.~4, 2015.

\bibitem{ahn2006dielectrophoretic}
K.~Ahn, C.~Kerbage, T.~P. Hunt, R.~Westervelt, D.~R. Link, and D.~A. Weitz, ``Dielectrophoretic manipulation of drops for high-speed microfluidic sorting devices,'' {\em Applied Physics Letters}, vol.~88, no.~2, 2006.

\bibitem{jing2015jetting}
T.~Jing, R.~Ramji, M.~E. Warkiani, J.~Han, C.~T. Lim, and C.-H. Chen, ``Jetting microfluidics with size-sorting capability for single-cell protease detection,'' {\em Biosensors and Bioelectronics}, vol.~66, pp.~19--23, 2015.

\bibitem{niu2007real}
X.~Niu, M.~Zhang, S.~Peng, W.~Wen, and P.~Sheng, ``Real-time detection, control, and sorting of microfluidic droplets,'' {\em Biomicrofluidics}, vol.~1, no.~4, 2007.

\bibitem{groisman2003microfluidic}
A.~Groisman, M.~Enzelberger, and S.~R. Quake, ``Microfluidic memory and control devices,'' {\em Science}, vol.~300, no.~5621, pp.~955--958, 2003.

\bibitem{gopinathan2023microfluidic}
K.~A. Gopinathan, A.~Mishra, B.~R. Mutlu, J.~F. Edd, and M.~Toner, ``A microfluidic transistor for automatic control of liquids,'' {\em Nature}, vol.~622, no.~7984, pp.~735--741, 2023.

\bibitem{martinez2024fluidic}
A.~Mart{\'\i}nez-Calvo, M.~D. Biviano, A.~H. Christensen, E.~Katifori, K.~H. Jensen, and M.~Ruiz-Garc{\'\i}a, ``The fluidic memristor as a collective phenomenon in elastohydrodynamic networks,'' {\em Nature Communications}, vol.~15, no.~1, p.~3121, 2024.

\bibitem{zeng20223d}
C.~Zeng, M.~W. Faaborg, A.~Sherif, M.~J. Falk, R.~Hajian, M.~Xiao, K.~Hartig, Y.~Bar-Sinai, M.~P. Brenner, and V.~N. Manoharan, ``3d-printed machines that manipulate microscopic objects using capillary forces,'' {\em Nature}, vol.~611, no.~7934, pp.~68--73, 2022.

\bibitem{PhysRevLett.118.198002}
R.~D. Schulman, R.~Ledesma-Alonso, T.~Salez, E.~Rapha\"el, and K.~Dalnoki-Veress, ``Liquid droplets act as ``compass needles'' for the stresses in a deformable membrane,'' {\em Phys. Rev. Lett.}, vol.~118, p.~198002, May 2017.

\bibitem{smith2021droplets}
K.~Smith-Mannschott, Q.~Xu, S.~Heyden, N.~Bain, J.~H. Snoeijer, E.~R. Dufresne, and R.~W. Style, ``Droplets sit and slide anisotropically on soft, stretched substrates,'' {\em Physical review letters}, vol.~126, no.~15, p.~158004, 2021.

\bibitem{ChaoPRL25}
Y.~Chao, H.~Jeon, and S.~Karpitschka, ``Nonmonotonic motion of sliding droplets on strained soft solids,'' {\em Phys. Rev. Lett.}, vol.~134, p.~184001, May 2025.

\bibitem{trujillo2008creasing}
V.~Trujillo, J.~Kim, and R.~C. Hayward, ``Creasing instability of surface-attached hydrogels,'' {\em Soft matter}, vol.~4, no.~3, pp.~564--569, 2008.

\bibitem{chen2014controlled}
D.~Chen, L.~Jin, Z.~Suo, and R.~C. Hayward, ``Controlled formation and disappearance of creases,'' {\em Materials Horizons}, vol.~1, no.~2, pp.~207--213, 2014.

\bibitem{dervaux2012mechanical}
J.~Dervaux and M.~B. Amar, ``Mechanical instabilities of gels,'' {\em Annu. Rev. Condens. Matter Phys.}, vol.~3, no.~1, pp.~311--332, 2012.

\bibitem{ciarletta2018matched}
P.~Ciarletta, ``Matched asymptotic solution for crease nucleation in soft solids,'' {\em Nature communications}, vol.~9, no.~1, p.~496, 2018.

\bibitem{cai2012creasing}
S.~Cai, D.~Chen, Z.~Suo, and R.~C. Hayward, ``Creasing instability of elastomer films,'' {\em Soft Matter}, vol.~8, no.~5, pp.~1301--1304, 2012.

\bibitem{CGS96}
A.~Carre, J.~Gastel, and M.~Shanahan, ``{Viscoelastic effects in the spreading of liquids},'' {\em {Nature}}, vol.~{379}, pp.~{432--434}, {FEB 1} {1996}.

\bibitem{AS2020}
B.~Andreotti and J.~H. Snoeijer, ``Statics and dynamics of soft wetting,'' {\em Annual Review of Fluid Mechanics}, vol.~52, no.~1, pp.~285--308, 2020.

\bibitem{Shanahan1994}
M.~E. Shanahan and A.~Carre, ``Anomalous spreading of liquid drops on an elastomeric surface,'' {\em Langmuir}, vol.~10, no.~6, pp.~1647--1649, 1994.

\bibitem{LALLang96}
D.~Long, A.~Ajdari, and L.~Leibler, ``Static and dynamic wetting properties of thin rubber films,'' {\em Langmuir}, vol.~12, no.~21, pp.~5221--5230, 1996.

\bibitem{karpitschka2015droplets}
S.~Karpitschka, S.~Das, M.~van Gorcum, H.~Perrin, B.~Andreotti, and J.~H. Snoeijer, ``Droplets move over viscoelastic substrates by surfing a ridge,'' {\em Nature communications}, vol.~6, no.~1, p.~7891, 2015.

\bibitem{pandey2018hydrogel}
A.~Pandey, C.~L. Nawijn, and J.~H. Snoeijer, ``Hydrogel menisci: Shape, interaction, and instability,'' {\em Europhysics Letters}, vol.~122, no.~3, p.~36006, 2018.

\bibitem{karpitschka2017cusp}
S.~Karpitschka, J.~Eggers, A.~Pandey, and J.~H. Snoeijer, ``Cusp-shaped elastic creases and furrows,'' {\em Physical review letters}, vol.~119, no.~19, p.~198001, 2017.

\bibitem{van2021pinning}
M.~A. Van~Limbeek, M.~H. Essink, A.~Pandey, J.~H. Snoeijer, and S.~Karpitschka, ``Pinning-induced folding-unfolding asymmetry in adhesive creases,'' {\em Physical review letters}, vol.~127, no.~2, p.~028001, 2021.

\bibitem{essink2023crease}
M.~H. Essink, M.~A. van Limbeek, A.~Pandey, S.~Karpitschka, and J.~H. Snoeijer, ``Adhesive creases: bifurcation, morphology and their (apparent) self-similarity,'' {\em Soft matter}, vol.~19, no.~27, pp.~5160--5168, 2023.

\bibitem{karpitschka2016liquid}
S.~Karpitschka, A.~Pandey, L.~A. Lubbers, J.~H. Weijs, L.~Botto, S.~Das, B.~Andreotti, and J.~H. Snoeijer, ``Liquid drops attract or repel by the inverted cheerios effect,'' {\em Proceedings of the National Academy of Sciences}, vol.~113, no.~27, pp.~7403--7407, 2016.

\bibitem{liu2019elastocapillary}
Q.~Liu, T.~Ouchi, L.~Jin, R.~Hayward, and Z.~Suo, ``Elastocapillary crease,'' {\em Physical review letters}, vol.~122, no.~9, p.~098003, 2019.

\bibitem{wilson1972excitatory}
H.~R. Wilson and J.~D. Cowan, ``Excitatory and inhibitory interactions in localized populations of model neurons,'' {\em Biophysical journal}, vol.~12, no.~1, pp.~1--24, 1972.

\bibitem{shockley1949theory}
W.~Shockley, ``The theory of p-n junctions in semiconductors and p-n junction transistors,'' {\em Bell system technical journal}, vol.~28, no.~3, pp.~435--489, 1949.

\bibitem{semenov2024preisach}
M.~Semenov, S.~Borzunov, P.~Meleshenko, and N.~Sel’Vesyuk, ``The preisach model of hysteresis: Fundamentals and applications,'' {\em Physica Scripta}, vol.~99, no.~6, p.~062008, 2024.

\bibitem{lindeman2025generalizing}
C.~W. Lindeman, T.~R. Jalowiec, and N.~C. Keim, ``Generalizing multiple memories from a single drive: The hysteron latch,'' {\em Science Advances}, vol.~11, no.~5, p.~eadr5933, 2025.

\bibitem{van2013domino}
J.~van Leeuwen, ``Domino magnification,'' {\em arXiv preprint arXiv:1301.0615}, 2013.

\bibitem{bilal2017bistable}
O.~R. Bilal, A.~Foehr, and C.~Daraio, ``Bistable metamaterial for switching and cascading elastic vibrations,'' {\em Proceedings of the National Academy of Sciences}, vol.~114, no.~18, pp.~4603--4606, 2017.

\bibitem{van2010domino}
J.~Van~Leeuwen, ``The domino effect,'' {\em American Journal of Physics}, vol.~78, no.~7, pp.~721--727, 2010.

\bibitem{mcarthur2023demonstration}
J.~A. McArthur, A.~A. Dadey, S.~D. March, A.~H. Jones, X.~Xue, R.~Salas, J.~C. Campbell, and S.~R. Bank, ``Demonstration of the alinassb cascaded multiplier avalanche photodiode,'' {\em Applied Physics Letters}, vol.~123, no.~4, 2023.

\bibitem{wang2020electrowetting}
H.~Wang and L.~Chen, ``Electrowetting-on-dielectric based economical digital microfluidic chip on flexible substrate by inkjet printing,'' {\em Micromachines}, vol.~11, no.~12, p.~1113, 2020.

\bibitem{paratore2022reconfigurable}
F.~Paratore, V.~Bacheva, M.~Bercovici, and G.~V. Kaigala, ``Reconfigurable microfluidics,'' {\em Nature Reviews Chemistry}, vol.~6, no.~1, pp.~70--80, 2022.

\bibitem{katsikis2015synchronous}
G.~Katsikis, J.~S. Cybulski, and M.~Prakash, ``Synchronous universal droplet logic and control,'' {\em Nature Physics}, vol.~11, no.~7, pp.~588--596, 2015.

\bibitem{sokuler2010softer}
M.~Sokuler, G.~K. Auernhammer, M.~Roth, C.~Liu, E.~Bonacurrso, and H.-J. Butt, ``The softer the better: fast condensation on soft surfaces,'' {\em Langmuir}, vol.~26, no.~3, pp.~1544--1547, 2010.

\bibitem{srinivasarao2001three}
M.~Srinivasarao, D.~Collings, A.~Philips, and S.~Patel, ``Three-dimensionally ordered array of air bubbles in a polymer film,'' {\em Science}, vol.~292, no.~5514, pp.~79--83, 2001.

\bibitem{stone2004engineering}
H.~A. Stone, A.~D. Stroock, and A.~Ajdari, ``Engineering flows in small devices: microfluidics toward a lab-on-a-chip,'' {\em Annu. Rev. Fluid Mech.}, vol.~36, no.~1, pp.~381--411, 2004.

\bibitem{prakash2008surface}
M.~Prakash, D.~Qu{\'e}r{\'e}, and J.~W. Bush, ``Surface tension transport of prey by feeding shorebirds: the capillary ratchet,'' {\em science}, vol.~320, no.~5878, pp.~931--934, 2008.

\bibitem{tuteja2008robust}
A.~Tuteja, W.~Choi, J.~M. Mabry, G.~H. McKinley, and R.~E. Cohen, ``Robust omniphobic surfaces,'' {\em Proceedings of the National Academy of Sciences}, vol.~105, no.~47, pp.~18200--18205, 2008.

\bibitem{seemann2011droplet}
R.~Seemann, M.~Brinkmann, T.~Pfohl, and S.~Herminghaus, ``Droplet based microfluidics,'' {\em Reports on progress in physics}, vol.~75, no.~1, p.~016601, 2011.

\bibitem{nosonovsky2009superhydrophobic}
M.~Nosonovsky and B.~Bhushan, ``Superhydrophobic surfaces and emerging applications: Non-adhesion, energy, green engineering,'' {\em Current Opinion in Colloid \& Interface Science}, vol.~14, no.~4, pp.~270--280, 2009.

\end{thebibliography}

\makeatletter
\let\old@makecaption\@makecaption
\renewcommand{\@makecaption}[2]{\old@makecaption{#1}{#2}}
\makeatother
\captionsetup{format=plain,justification=justified,singlelinecheck=false}

\end{document}